\documentclass[letters]{aa} 

\usepackage{graphicx}
\usepackage{txfonts}
\begin{document}

\title{Using Observations of Distant Quasars to Constrain Quantum Gravity}

\author{Eric S. Perlman
	\inst{1}
	\and
	Y. Jack Ng
	\inst{2}
	\and
	David J. E. Floyd
	\inst{3}
	\and
	Wayne A. Christiansen
	\inst{2}
	}
	
\institute{Department of Physics \& Space Sciences, Florida Institute of 
	Technology, 150 W. University Blvd., Melbourne, FL  32901, USA \\
	\email{eperlman@fit.edu}
	\and
	Department of Physics and Astronomy, University of North Carolina,
	Chapel Hill, NC  27599, USA \\
	\email{yjng@physics.unc.edu,wayne@physics.unc.edu}
	\and
	Monash Centre for Astrophysics, School of Physics, Monash University, 
	P. O. Box 27, Clayton, Victoria 3800, Australia \\
	\email{david.floyd@monash.edu} 	
	}
\date{}

\abstract{}
 	{The small-scale nature of spacetime can be tested with observations of 
	distant quasars.  We comment on a recent paper by Tamburini et al. (A\&A, 533, 71) 
	which claims that Hubble Space Telescope ({\it HST}) observations of the 
	most distant quasars place severe constraints on models of foamy 
	spacetime. }
   	{ If space is foamy on the Planck scale, photons emitted from distant 
	objects will accumulate uncertainties in distance and propagation 
	directions thus affecting the expected angular size of a compact 
	object as a function of redshift.  We discuss the geometry of 
	foamy spacetime, and the appropriate distance measure for calculating 
	the expected angular broadening.  We also address the mechanics of
	carrying out such a test.  
	We draw upon our previously published work on this subject, 
	which carried out similar tests as Tamburini et al.
	and also went considerably beyond their work in several 
	respects.  }
   	{When calculating the path taken by photons as they travel from a
	distant source to Earth, one must use the comoving distance 
	rather than the luminosity distance.  This then also becomes the 
	appropriate distance to use when calculating the angular broadening
	expected in a distant source.  The use of the wrong distance measure 
	causes Tamburini et al. to overstate the constraints that can be placed 
	on models of spacetime foam.  In addition, we consider the impact of 
	different ways of parametrizing and measuring the effects of spacetime 
	foam.  Given the variation of the shape of the point-spread 
	function (PSF) on the chip, as well as 
	observation-specific factors, it is important to 
	select carefully -- and 
	document -- the comparison stars used as well as the methods used
	to compute the Strehl ratio.}
{}

	
	\maketitle
	

\section{Introduction}

Even at the minute scales of distance and duration examined with
increasingly discriminating instruments, spacetime still appears
to be smooth and structureless. However, a variety of models of 
quantum gravity posit that spacetime is, on Planck scales, subject to
quantum fluctuations.  Hence, if probed at a small enough scale, spacetime 
will appear complicated -- something akin in complexity to a turbulent
froth that ~\cite{whe63}
has dubbed ``quantum foam,'' also known as ``spacetime
foam.''
The detection of spacetime foam is important for constraining models of 
quantum gravity.  If a foamy structure is found, it would require 
at least a probabilistic rather than deterministic nature of 
spacetime itself, as the paths taken by different photons emitted by a 
distant source would not be identical to one another.
   
In this commentary paper, we discuss the use of astronomical observations of 
distant sources to test models of quantum gravity.  We concentrate particularly
on a recent paper by \cite{tam}, published in Astronomy \& Astrophysics in 
September 2011.  Some of the points discussed below were discussed in our own
paper \cite{CNFP},
which was published nine months earlier. 
The present paper is organized as follows.  In \S 2 we discuss the nature 
of quantum
fluctuations and the proper distance measure to use.  This has important
implications for the predicted size of the seeing disk, and hence the 
constraints one can put on spacetime foam models given a
non-detection, as we then discuss in \S 3. 
In \S 4 we discuss practicalities of carrying out these tests.  These include
the need to characterize the point-spread function (PSF) of a given telescope
in terms that can be compared to the profile observed in a distant, unresolved
source, such as a quasar or supernova.  
Finally, in \S 5 we close with a summary.


\section{The Nature of Quantum Fluctuations and the Predicted Seeing Disk}
	  
To quantify the problem, let us recall that, if spacetime 
undergoes quantum fluctuations, the intrinsic distance to an object will vary,
thus producing an intrinsic limitation to the accuracy with which one can 
measure a macroscopic distance.  If we denote 
the fluctuation of a distance $l$ by $\delta l$,  
we expect $\delta l \gtrsim N l^{1 - \alpha} l_P^{\alpha}$,
(see \cite{ng03b}), where $N$ is a numerical factor $\sim 1$ and 
$l_P = \sqrt{\hbar G/c^3}$ is the Planck length,
the characteristic length scale in quantum gravity.
The length in this expression, $\delta l$, must be defined with reference to 
the macroscopic distance, $l$ (rather than locally).
The parameter $\alpha
\lesssim 1$ specifies the different spacetime foam models.

Distance fluctuations $\pm \delta l$ imply phase fluctuations
$\pm \Delta \phi = \pm 2 \pi \delta l / \lambda$ 
(see ~\cite{lie03,rag03,ng03a}).  
One practical method of searching for these fluctuations is to look for 
``halos'' in images of distant, unresolved sources, which can be produced by 
fluctuations in the direction of the local wave-vector, 
$\pm \delta \psi \equiv
\pm \Delta \phi /(2 \pi) = \pm \delta l / \lambda$.
The point is that  due to
quantum foam-induced fluctuations in the phase velocity of an incoming light
wave from a distant point source, the wave front itself develops a small
scale ``cloud of uncertainty'' equivalent to a ``foamy'' structure, 
because parts of the wave-front lag while other
parts advance. 
This results in the wave vector, upon detection, acquiring a jitter in
direction with an angular spread of the order of
$\delta \psi$. 
In effect, spacetime foam creates a ``seeing disk'' whose angular diameter is
\begin{equation}
\delta \psi
= N \big(\frac {l}{\lambda}\big)^{1 - \alpha} \big(\frac {l_P}{\lambda}\big)^{\alpha}. 
\label{eq0}
\end{equation}
We note that the magnitude of $\delta \psi$ as given in the above 
equation is consistent with our assumption of isotropic fluctuations
which implies comparable sizes of the wave-vector fluctuations
perpendicular to and along the line of sight (see \cite{chr06}).
For a telescope or
interferometer with baseline length $D_{tel}$, this
means that the dispersion ($\sim \delta \psi$, normal
to the wave front) will be recorded as a spread in the angular size of a
distant point source, causing a reduction in the Strehl ratio, and/or the
fringe visibility when
$\delta \psi \sim \lambda / D_{tel}$ for a diffraction limited telescope.

The fundamental uncertainties caused by
spacetime foam are spatial, not angular, even though they result in a ''seeing
disk''.  Strictly speaking, the models specify
the uncertainty $\pm \delta l$, in distance between a source and observer along
the line of sight.  This is because $\delta l$ is defined by the uncertainty in
the distance measured by light travel times.  Of course, there is also a 
corresponding
uncertainty in the transit time for light from source to observer,  $\delta t
\sim \delta l/c$.   Furthermore, since the
globally averaged wavefront is effectively spherical, globally averaged 
photon trajectories will deviate from the direct line of sight by an angle
less than or equal to $\delta l/l$.  As a direct consequence, the 
expected blurring of distant images is {\it not}
the result of a random walk of small angle photon scatterings along the line of
sight, since the uncertainties in the derived directions of the local wave
vectors must result in the same spatial uncertainty, $\delta l$ (no matter how
many wave crests pass the observer's location).  For example, in the "thin
screen approximation", the accumulated transverse path of multiply scattered
photons would be approximated as $(\delta \psi)l >> \delta l$.  This 
would lead to
expected time lags, $\delta \psi (l/c) >> \delta l/c$, in conflict with the
basic premises for spacetime foam models.  

The above background, given in greater detail in our recent paper 
\cite{CNFP}, 
illustrates why, when measuring the length $l$ for sources at cosmological
distances, the appropriate distance measure to use is 
the line-of-sight comoving distance (see ~\cite{hog00}) given by
\begin{equation}
D_C(z) = D_H I_E(z)
\label{eq1}
\end{equation}
where
\begin{equation}
I_E (z)= \int_0^{z} {\frac{dz'}{E(z')}},
\end{equation}
and
\begin{equation}
E(z) = \sqrt{\Omega_M (1+z)^3 + \Omega_k (1 + z)^2 + \Omega_{\Lambda}},
\label{eq2}
\end{equation}
with $D_H = c/H_0$ being the Hubble distance, $\Omega_M, \Omega_k$ and
$\Omega_{\Lambda}$ being the (fractional) density parameter associated with
matter, curvature and the cosmological constant respectively. Consistent
with the latest WMAP + CMB data, we will use
$\Omega_M = 0.25, \Omega_{\Lambda} = 0.75$ and
$\Omega_k = 0$, and for the Hubble distance we will use
$D_H = 1.3 \times 10^{26}$ meters.

\section{Predicting the Halo Size}

In terms of the comoving distance, for the various models of spacetime foam
(parametrized by $\alpha$), the equivalent halo size is given by
\begin{equation}
\delta \psi = {\frac{N (1 - \alpha) l_P^{\alpha} D_H^{1 - \alpha}
I(z,\alpha)}{\lambda_o}}, 
\label{eq3}
\end{equation}
with
\begin{equation}
I(z,\alpha) = \int_0^z {\frac {dz' (1+z')}{E(z')}} I_E(z')^{-\alpha},
\label{eq4}
\end{equation}
where the factor $(1 + z')$ in the integral corrects the observed wavelength
$\lambda_o$, back to the wavelength
$\lambda (z')$ at redshift $z'$. That is,
$\lambda (z') = \lambda_o/(1 + z')$.

\begin{figure}
\centering
{\includegraphics[width=9cm]{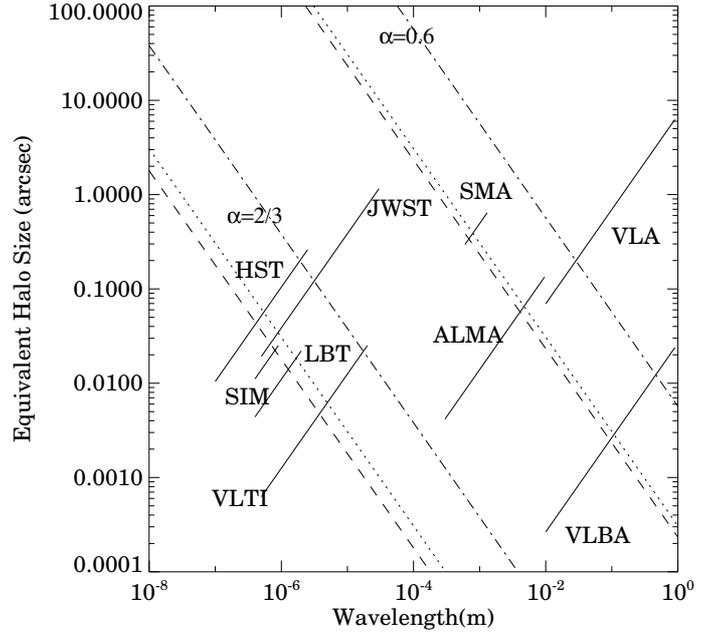}}
\caption{\label{fig:det} The detectability of various models of foamy spacetime 
for existing and planned telescopes.  We show the diagonal tracks for 
halo size $\delta \psi$ for an unresolved, $z=6.3$ source, using
the comoving distance [eq. (5), dashed lines], naive application of the luminosity
distance [i.e., not redoing the integral $I(z,\alpha)$ as per equation (9), 
in dotted lines], and correct application of the luminosity distance (dash-dot
lines).  Tracks are shown for $\alpha=0.6, 2/3$, and $N=1.8$. 
See \S\S 2,3 for discussion.  It appears to us that \cite{tam} used the 
phase uncertainty $\Delta \phi = 2 \pi \delta \psi$ as a measure of halo 
size, which would exaggerate the expected halo size by nearly an order of 
magnitude.  This displacement would make it appear that quantum
foam may be easily tested by HST imaging, which it is not.} 
\end{figure}

We have used these results to produce Figure 1.  The diagonal lines in Figure 1
show predictions for the size of the seeing disk for different models of
spacetime foam, for a source at redshift $z=6.3$, which represents the highest 
redshift
quasar examined by \cite{tam}.  We note that $\delta \psi$ in Figure 1 is 
a factor of $2 \pi$ smaller than the phase, $\Delta \phi$, which was used
incorrectly by \cite{tam} to calculate expected halo size.
In the case of a non-detection of angular 
broadening, the
region above the diagonal line for a given $\alpha$ may be excluded.  

The discussion above illustrates the importance of a correct understanding of
the seeing disk caused by spacetime foam as a spatial, rather than angular
effect, thus requiring the use of the comoving distance.  
Figure 1
also shows how the prediction (for $\delta \psi$ not $\Delta \phi$)
changes if one were to incorrectly model the
seeing disk as being the result of angular fluctuations, and hence use the
luminosity distance,
\begin{equation}
D_L(z)=(1+z)D_C(z) = (1+z)D_H I_E(z),
\end{equation} 
rather than the comoving distance.   
This is the assumption
made by \cite{tam} as well as  \cite{ste07}.

As an illustration of the cosmological effects we use equation (7) to
calculate the equivalent halo size that one would predict {\it if one 
incorrectly} used the luminosity distance.
To do this we make use
of the last part of equation (7) as our $l'$, in which case 
\begin{equation}
dl'= dD_L(z') = dz' D_H\int_0^{z'} {\frac{dz''}{E(z'')}} + {\frac{(1+z')D_H
dz'}{E(z')}}.
\end{equation}       
The result is that in calculating $\delta\psi$ one cannot simply 
use the luminosity distance in equation (5), 
and multiply it by $(1+z)$.  Instead one
must replace $I(z,\alpha)$ in equation (6) with the following:
\begin{equation}
I_2(z,\alpha)=\int_0^z dz'(1+z')\big[{{(1+z')}\over{E(z')}}+I_E(z')\big]\big[(1+z')I_E(z')\big]^{-\alpha}.
\end{equation}
 
Unfortunately, \cite{tam} do precisely this (their equations (2) and (3)).  We
can use the above formalism to estimate how this affects their quoted
constraints.  In Figure 1 we have overplotted tracks for the application of
luminosity distance, both  in the case of redoing $I(z,\alpha)$ and not redoing
the integral.  
As can be seen, in Figure 1 the use of the incorrect distance measure causes a
rather large miscalculation of the expected halo size that leads to an 
exaggeration in size by a factor of about 20 at a given wavelength. 
Furthermore, because the parallel set of diagonal lines in Figure 1 represents
trajectories for $\delta\psi$ versus $\lambda$ that are specified by 
$\alpha$, this reduction in halo size leads to a reduction
in the limiting value of $\alpha$ that can be determined from observations.  
\cite{tam} claim that current data exclude 
models with $\alpha < 0.68$ ($a_0 \sim 1$, 
including the holographic model which has $\alpha=2/3$) 
(the ``red zone'' in their Figure 5).  
However, by using the correct co-moving distance, we find that
their limit for excluding quantum foam models 
should be reduced by $\Delta \alpha = 0.021$, much 
more consistent with the limit of   
$\alpha \sim 0.65$ previously established by \cite{CNFP}.


\section{Observing Practicalities:  Strehl Ratio and PSF} 

In conventional imaging 
the best way to characterize the halo is in terms of the observed
Strehl ratio.  This is defined as the ratio of the observed peak intensity from
a point source as compared to the theoretical maximum peak intensity of a 
perfect telescope working at its diffraction limit.  
As can be seen by reference to Fig. 1, quasars are expected to
be barely resolved in {\it HST} observations, and the Strehl ratio gives 
a concrete way to quantify
how unresolved they are.  This comparison must be done with reference to known
stars in one's image, because the PSF of the {\it HST}
varies significantly with position on the focal plane (and hence each individual
camera).  The Strehl ratio is defined as the ratio of the observed image peak to
the peak diffraction spike.  In \cite{CNFP} we approximated this ratio as
\begin{equation}
S_\mathrm{Obs} = S_M \exp \left[-(\sigma_I^2 + \sigma_{\psi}^2)\right]
\label{eq5}
\end{equation}
where $S_M \leq 1$ represents a degradation of the observed Strehl ratio
due to masking effects,
$\sigma_I$ represents uncorrelated wavefront errors induced by the instrumentation
(i.e., telescope plus instruments) and $\sigma_{\psi}$ represents uncorrelated 
wavefront
``errors'' induced by spacetime foam. Both of these
dispersions are expressed in units of the telescope's diffraction limit, 
$\lambda / D_{tel}$. 
A similar treatment is taken in \cite{tam}, along with a superficially similar 
procedure, although it should be noted that they do not publish a list of the 
comparison stars used to compute $\sigma_I$ (as we did explicitly in \cite{CNFP}).
This last makes it difficult to reproduce their results.
 
If we follow this prescription, we can then define the spacetime foam degraded
Strehl $S_{\psi}$ as $S_{\psi} = \exp (-\sigma_{\psi}^2)$,
where $\sigma_{\psi}$ is $\delta \psi$ divided by 
$\lambda / D_{tel}$.   
Provided the comoving distance is used, as argued in \S 2, 
we then obtain for $\sigma_{\psi}$ 
\begin{equation}
\sigma_{\psi} = {\frac{N (1 - \alpha) l_P^{\alpha} D_H^{1 - \alpha} I(z,
\alpha) D_{tel}}{\lambda_o^2}}.
\label{sigma}
\end{equation}
This approximation, of course, breaks down when $\sigma_{\psi} \sim 1$, i.e.,
when the wave front angular dispersion is comparable to the telescope's angular
resolution. A fully parametrized version of the resultant Strehl ratio then is
\begin{equation}
S_{\psi} = \exp \left({\frac{- N^2 (1 - \alpha)^2 l_P^{2\alpha} D_H^{2(1 - \alpha)} 
I^2 (z,\alpha) D_{tel}^2}{ \lambda_o^4}} \right).
\label{eq:SR}
\end{equation}

\begin{figure}
{\centering{
\includegraphics[width=9cm]{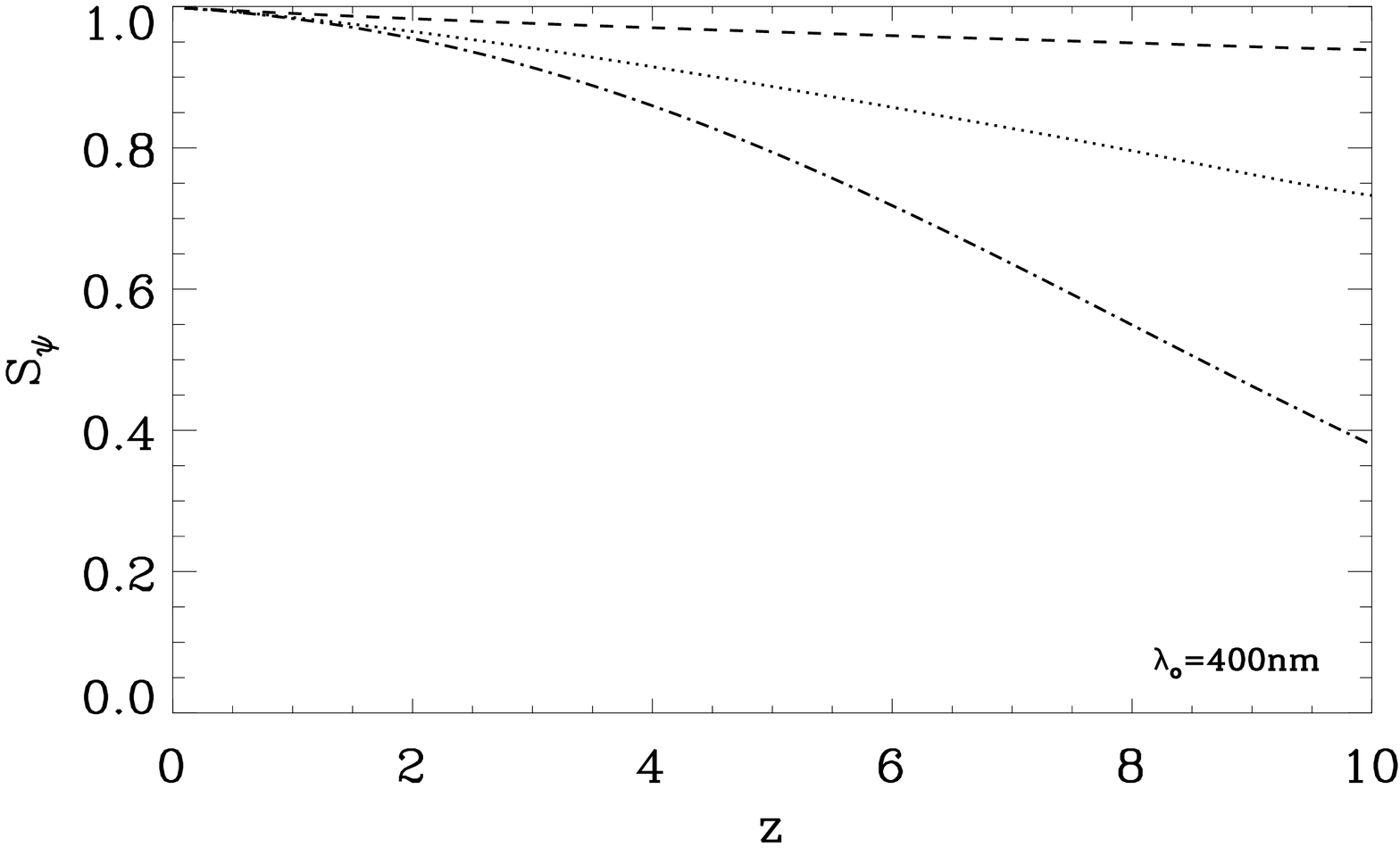}
\includegraphics[width=9cm]{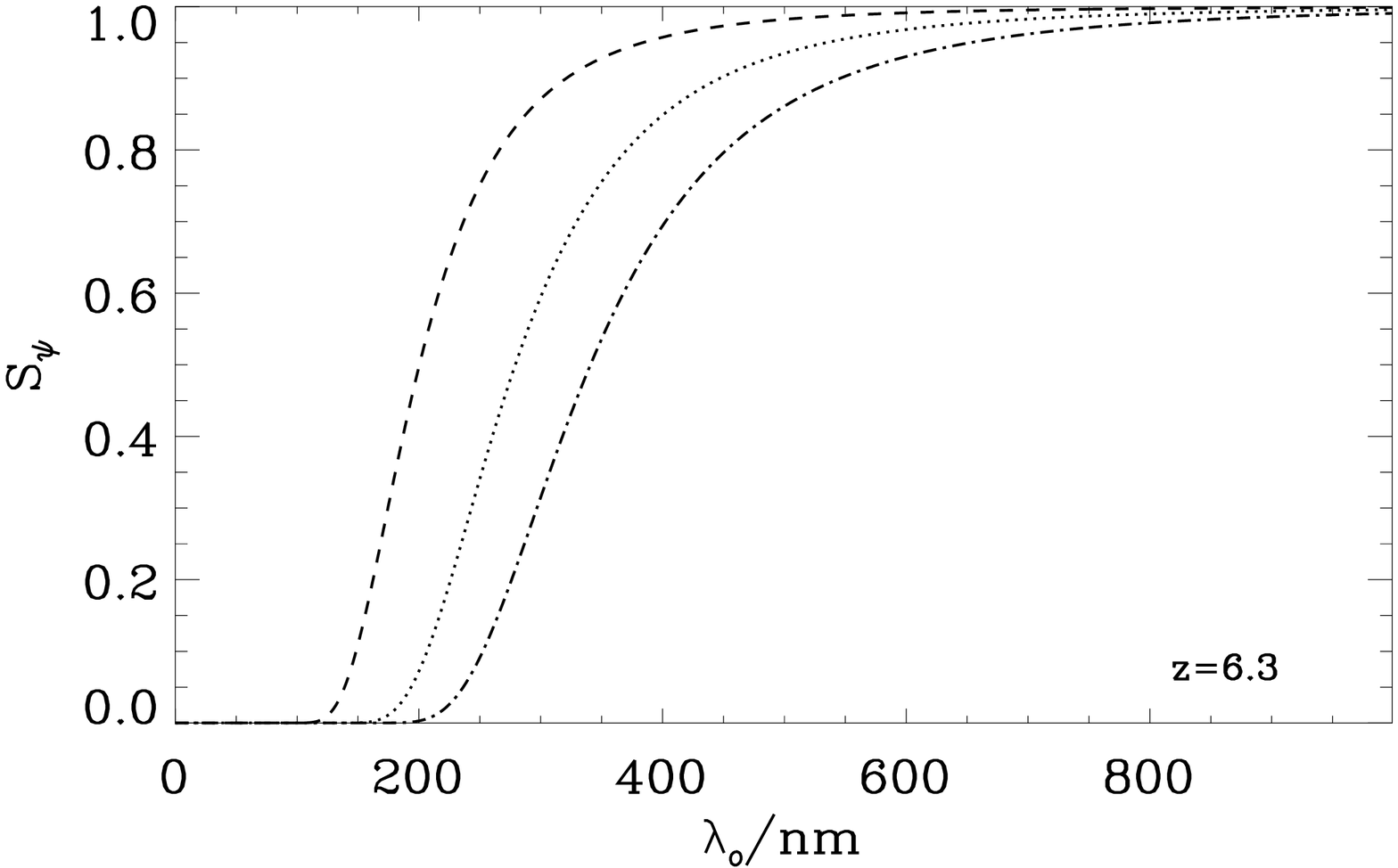}}}
\caption{Expected Strehl ratio as a function of redshift (top) and wavelength
(bottom).  In both plots, we assume $\alpha=2/3$ and $N=1.8$.  In the top 
figure
we examine the change in Strehl ratio expected for a point source with varying
redshift, $z$, for an observed wavelength of 400 nm.  
In the bottom 
plot, we specifically show the case of a $z=6.3$ quasar, as examined by
\cite{tam}, for an observed wavelength of 400 nm.  
As in Figure 1, the dashed line refers to the use of comoving 
distance, dotted line refers to the naive use of luminosity distance, and
dash-dot line refers to the treatment in equation (9).}
 \end{figure}

However, just as with the expected halo size, the use of the luminosity
distance  drastically affects this expression.  We cannot simply replace $D_C$
in equation (12) by $D_L$, as was done in \cite{tam}.  Instead, 
$I(z,\alpha)$ must also be replaced 
with $I_2(z,\alpha)$ (equation (9)).  This causes an overestimate
in the magnitude of the exponential argument, thus causing a corresponding 
reduction in the Strehl ratio which is consistent with the discussion following
equation (9).  

In Figure 2, we show the result of this error. 
As can be seen, even in the case of a source at $z=6.3$ -- the highest redshift
source considered by \cite{tam} -- the effects of spacetime foam simply are not
detectable in {\it HST} observations.  From Figure 2, as was pointed out in
\cite{CNFP}, it is not surprising that effects of spacetime foam are likely not
to be detectable in HST images cross-referenced with high redshift SDSS quasars,
because the only Hubble images from the SDSS  sample are in the near IR band. At
a wavelength of 8000 \AA, typical of the observations used in \cite{tam},  which
used the ACS + F775W and F850LP filters, the expected Strehl ratio is 
$S_{\psi}> 0.98$
for  both the comoving distance as well as a naive application of the luminosity
distance -- i.e., just using eq. (7) and not including the modified  integral
$I_2$ (equation (9)). Even if both of these factors are included,  we  
still expect $S_\psi > 0.95$, at most just a 5\%
reduction in the measured Strehl with respect to that of the instrument. 
By comparison, in our paper
(\cite{CNFP}, Table III), we used comparison  stars for  the HUDF quasars to
measure the {\em instrumental} Strehl ratios, finding values  between $S_I=0.27$
(F435W) and 0.64 (F850LP), with the low Strehl ratios in the  blue being a
result of the undersampling of the PSF by the ACS.   While we are unable to
comment exactly on the success of the method of \cite{tam} because they did not
specify which stars were used or provide adequate information on the mechanics
of deriving the phase (in their formulation), we can say that we find highly
unrealistic  their claim to have achieved the maximum possible constraint on
$\alpha$ for this wavelength, based on our extensive experience  with {\it HST}
data.   
Indeed, as our  work showed (\cite{CNFP}, Table
IV) even for the much deeper observations of the UDF  quasars (which extended to
shorter wavelengths,  but were at typical redshifts $z=4$, translating to a
comoving  distance about 15\% lower than $z=6.3$), the corrected Strehl ratio,
$S_M / S_I$, that was achieved ranged from 1.04 down to 0.40, depending on the
band, with two of the four  being at Strehl ratios of $\sim 0.90$.   
The lower Strehl ratios 
were no doubt caused by a combination of factors, including not only the
imperfections in the PSF of the {\it HST}  and distortions across the chip and 
light path of individual images, but also factors intrinsic to the QSO
such as the host galaxy.  This is why in our paper, even though theoretically 
the observations of the HUDF quasar could probe to $\alpha \sim 0.66$, in
practice the constraint that could be set was only $\alpha=0.65$ (see Figure 5
in \cite{CNFP}).  On the basis of our experience, we believe it is likely that a
similar statement can be made for the observations examined by \cite{tam}.

It is worth mentioning that with current telescopes a second method of measuring
possible effects of spacetime foam is becoming available.  This is through the
use of interferometry, {\it e.g.,} by using the VLTI.  As can be seen in Figure
1, the VLTI would have a significant advantage in resolution over any optical-IR
telescope, simply because its longest baseline is $\sim$ factor 20 longer than 
the largest telescope currently in use or under construction.  Moreover, it
would not suffer from some of the problems we have noted in the {\it HST}
observations, namely undersampling.  As well, since interferometers are very
effective spatial filters the effect of the quasar's host galaxy would also be
minimized.  We therefore believe that the best way to probe the $\alpha\sim 0.7$
regime is with interferometers. 

We should point out that time lags from distant pulsed sources have also been
posited as a possible test of quantum foam models.  But, as
explained in \cite{CNFP},  {\it the new Fermi Gamma-ray Space Telescope results
(\cite{Abdo}) only exclude models with $\alpha < 0.3$.}

\section{Summary}

We have reviewed the theoretical basis for expecting halos due to spacetime
foam, and also the correct distance measure.  We have shown explicitly that,
while we agree with the basic result of \cite{tam} that current observations
with {\it HST} show no evidence for quantum gravity, as shown in our previously 
published paper \cite{CNFP}, we cannot agree with the resulting constraint they
placed on models of quantum gravity.  Because their calculations overstated the
size of quantum foam induced halos of distant quasars by a factor 20, their 
limit for $\alpha$, is also overstated by a minimum of
$\Delta\alpha=0.021$.  Based on our experience with {\it HST} data, we also 
believe -- but cannot verify (because of the lack of documentation in
\cite{tam}) -- that even this resulting level ($\alpha=0.66$) cannot be reached
because of details specific to each observation, including the variation of the 
PSF of {\it HST} with position on the chip, the undersampling of the PSF by
every instrument on {\it HST}, as well as the host galaxy of the
quasar.

\bigskip

{\noindent This work was supported in part by the US Department of Energy under
contract DE-FG02-06ER41418.}

\end{document}